\documentclass[paper]{bib/ieice}
\usepackage{graphicx}
\usepackage{latexsym}
\usepackage[fleqn]{amsmath}
\usepackage[psamsfonts]{amssymb}
\usepackage{amsthm}
\usepackage{color}

\def\x{\boldsymbol{x}}

\def\D{\boldsymbol{D}}

\def\X{\boldsymbol{X}}

\def\0{\boldsymbol{0}}
\def\cA{\mathcal A}

\def\cI{\mathcal I}

\def\cP{\mathcal P}

\def\cS{\mathcal S}

\def\cU{\mathcal U}

\def\cX{\mathcal X}
\def\cY{\mathcal Y}
\def\cZ{\mathcal Z}

\def\ov{\overline}

\def\wh{\widehat}
\def\wt{\widetilde}

\def\eqdef{{\displaystyle\mathop{=}^{\mbox{\rm def.}}}}

\newcommand\dotdot[1]{\mbox{$\ddot{\mbox{#1}}$}}
\newcommand\dash[1]{\mbox{$\acute{\mbox{#1}}$}}%

\newtheorem{teigi}{Definition}
\newtheorem{teiri}{Theorem}

\newtheorem{hodai}{Lemma}

\field{A}
\title{%
  Multiterminal source coding with complementary delivery
}
\titlenote{
  Some of the material in this paper was presented at the 2006 International Symposium on
  Information Theory and Its Applications, Seoul, Korea October-November 2006.
}
\authorlist{%
 \authorentry[research at akisato org]{Akisato Kimura}{m}{cslab}
 \authorentry[uyematsu at ieee org]{Tomohiko Uyematsu}{m}{titech}
}
\affiliate[cslab]{
 NTT Communication Science Laboratories, NTT Corporation,
 3-1 Morinosato Wakamiya, Atsugi-shi, Kanagawa, 243-0198 Japan.
}
\affiliate[titech]{%
 Department of Communications and Integrated Systems, Tokyo Institute of Technology,
 2-12-1 Ookayama, Meguro-ku, Tokyo, 152-8552 Japan.
}
\received{2008}{3}{23}

\begin{document}
\allowdisplaybreaks  
\maketitle
\begin{summary}
A coding problem for correlated information sources is investigated. Messages emitted
from two correlated sources are jointly encoded, and delivered to two decoders. Each
decoder has access to one of the two messages to enable it to reproduce the other
message. The rate-distortion function for the coding problem and its interesting
properties are clarified.
\end{summary}

%
\section{Introduction}

Coding problems for correlated information sources were originally investigated by
Slepian and Wolf \cite{SlepianWolf}. Corresponding rate-distortion coding problems
\cite{WynerZiv,MultiterminalSourceCoding:Longo,PhD:tung} and various coding problems
(e.g. \cite{SideInformationCoding:Wyner,SgarroCoding,KornerMarton}) inspired by the work
by Slepian and Wolf have been considered. Including the above studies, the main focus
in the 1970's was on coding problems with {\it separate encoding} (each message is
separately encoded) and {\it joint decoding} (several codewords are sent to a decoder
and decoded simultaneously).

In contrast, since the 1980's, coding problems that involve {\it joint encoding}
(messages from several sources are encoded at once) and/or {\it separate decoding} (each
message is separately decoded) have been explored. Separate decoding processes have
mainly been considered in relation to multiple description (e.g.
\cite{MultipleDescription:witsenhausen,MultipleDescription:wolf,%
MultipleDescription:ElGamalCover}), while joint encoding processes can be seen, for
example, in the cascading and branching communication systems \cite{CascadeAndBranch}, the
triangular communication system \cite{TriangularYamamoto} and multi-hop networks
\cite{MultipleNetworkDCC,MultipleNetworkISIT}.

\begin{figure}[t]
  \begin{center}
    \includegraphics[width=8cm]{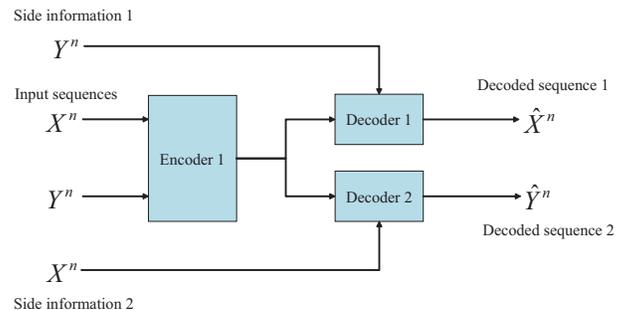}
    \caption{Complementary delivery network}
    \label{fig:diagram_full}
  \end{center}
\end{figure}

Also, a coding problem that involves joint encoding and separate decoding was considered
by Willems et al. \cite{BroadcastSatelliteCodingOrig,BroadcastSatelliteCoding}. The
coding system models a communication network via a satellite. Several stations are
separately deployed in a field. Every station collects its own target data and wants to
share all the target data with the other stations. To accomplish this task, each station
transmits the collected data to a satellite, and the satellite broadcasts all the
received data back to the stations. Each station utilizes its own target data as side
information to reproduce all the other target data. Willems et al.
\cite{BroadcastSatelliteCoding} investigated a special case of the above scenario in
which three stations were deployed and each station had access to one of three target
messages, and determined the minimum {\it lossless} achievable rate for uplink (from
each station to the satellite) and downlink (from the satellite to all the stations)
transmissions. Their main result implies that the uplink transmission is equivalent to
the traditional Slepian-Wolf coding system \cite{SlepianWolf}, and thus the main problem
is the downlink part. Henceforth we denote the networks characterized by the downlink
transmission as {\it generalized complementary delivery networks}, and we denote the
generalized complementary network with two stations and two target messages as the
{\it complementary delivery network} (Fig. \ref{fig:diagram_full}). This notation is
based on the network structure where each station (i.e. decoder) complements the target
messages from the codeword delivered by the satellite (i.e. encoder). Kimura et al.
investigated a universal coding problem for the complementary delivery network
\cite{UniversalComplementaryPaper} and the generalized complementary delivery network
\cite{GeneralizedUniversalComplementaryPaper}, and proposed an explicit construction of
lossless universal codes which attains the optimal error exponent. Also, Kuzuoka et al.
\cite{LinearUniversalComplementarySTW,LinearUniversalComplementaryPaper} simplified the
coding scheme by introducing a concept of network coding \cite{LinearNetworkCoding}.

The above previous researches considered only the lossless coding problem. In contrast,
this paper focuses on the lossy coding problem. The minimum achievable rate given
distortion criteria and some interesting properties of the minimum achievable rate are
clarified.

This paper is organized as follows. Section \ref{sec:define} provides notations and
definitions used throughout in this paper. Section \ref{sec:full} investigates the lossy
coding problem for the complementary delivery network, which includes descriptions of the
main result and several related properties. The main result can be easily extended to the
problem of the generalized complementary delivery networks, which will be discussed in
Section \ref{sec:many}. Finally, Section \ref{sec:proof} provides theorem proofs.

\section{Preliminaries}
\label{sec:define}

Let $\cX$ and $\cY$ be finite sets. Especially, for any natural number $M$, we denote
$\cI_M=\{1,2,\cdots,M\}$. The cardinality of $\cX$ is denoted as $|\cX|$. A member of
$\cX^n$ is written as $x^n=(x_1,x_2,\cdots,x_n)$, and substrings of $x^n$ are written as
$x_i^j=(x_i,x_{i+1},\cdots,x_j)$ for $i\le j$. A set of all the probability distributions
on $\cX$ is denoted as $\cP(\cX)$. A discrete memoryless source $(\cX,P_X)$ is an
infinite sequence $\{X_i\}_{i=1}^{\infty}$ of independent copies of a random variable
$X$ taking values in $\cX$ with a generic distribution $P_X\in\cP(\cX)$, namely
\begin{eqnarray*}
  P_{X^n}(x^n) &=& \prod_{i=1}^n P_X(x_i).
\end{eqnarray*}
$\cP(\cX|P_Y)$ denotes a set of all the probability distributions on $\cX$ given a
distribution $P_Y\in\cP(\cY)$. Namely, each member of $\cP(\cX|P_Y)$ is characterized by
$P_{XY}\in\cP(\cX\times\cY)$ as $P_{XY}=P_{X|Y}P_Y$. A source $(\cX,P_{X})$ can be
denoted by referring to its generic distribution $P_X$ or random variable $X$. For
a correlated source $(X,Y)$, $H(X)$, $H(X|Y)$ and $I(X;Y)$ denote the entropy of $X$, the
conditional entropy of $X$ given $Y$, and the mutual information of $X$ and $Y$,
respectively. Similarly, for a correlated source $(X,Y,Z)$, $I(X;Y|Z)$ denotes the
conditional mutual information of $X$ and $Y$ given $Z$. In the following, all bases of
exponentials and logarithms are set at $e$ (the base of the natural logarithm). Let
$\wh{\cX}$ stand for a reconstruction alphabet that corresponds to a source $X$ to be
encoded, and let $\Delta_X:\cX\times\wh{\cX}\to[0,\ov{\Delta}_X]$ be a corresponding
single-letter distortion function, where $\ov{\Delta}_X<\infty$. The vector distortion
function is defined in the usual way, i.e.
\begin{eqnarray*}
  \Delta_X^n(x^n,\wh{x}^n) &=& \frac 1n\sum_{k=1}^n\Delta_X(x_k,\wh{x}_k).
\end{eqnarray*}

\section{Complementary delivery}
\label{sec:full}
\subsection{Problem formulation}
\label{sec:full:formulate}

\begin{teigi}  \label{def:lossy:code}
  {\rm(CD (Complementary Delivery) code)}\\
  A set $(\varphi_n,\wh{\varphi}_n^{(1)},\wh{\varphi}_n^{(2)})$ of an encoder and
  decoders is a CD code $(n,M_n,\rho_n^{(X)},\rho_n^{(Y)})$ for the source $(X,Y)$
  if and only if
  \begin{eqnarray*}
    \varphi_n            &:& \cX^n\times\cY^n\rightarrow\cI_{M_n}\\
    \wh{\varphi}_n^{(1)} &:& \cI_{M_n}\times\cY^n\rightarrow\wh{\cX}^n,\\
    \wh{\varphi}_n^{(2)} &:& \cI_{M_n}\times\cX^n\rightarrow\wh{\cY}^n,\\
    \rho_n^{(X)} &=& E\left[\Delta_X^n(X^n,
                            \wh{\varphi}_n^{(1)}(A_n,Y^n))\right],\\
    \rho_n^{(Y)} &=& E\left[\Delta_Y^n(Y^n,
                            \wh{\varphi}_n^{(2)}(A_n,X^n))\right],\\
    A_n &=& \varphi_n(X^n,Y^n).
  \end{eqnarray*}
\end{teigi}

\begin{teigi}  \label{def:lossy:rate}
  {\rm(Lossy CD-achievable rate)}\\
  $R$ is a lossy CD-achievable rate of the source $(X,Y)$ for a given distortion pair
  $(D_X,D_Y)$ if and only if there exists a sequence
  $\left\{\left(n,M_n,\rho_n^{(X)},\rho_n^{(Y)}\right)\right\}_{n=1}^{\infty}$ of CD
  codes for the source $(X,Y)$ such that
  \begin{eqnarray*}
    && \limsup_{n\to\infty}\frac 1n\log M_n \le R,\\
    && \limsup_{n\to\infty}\rho_n^{(X)} \le D_X,~
       \limsup_{n\to\infty}\rho_n^{(Y)} \le D_Y.
  \end{eqnarray*}
\end{teigi}

\begin{teigi}  \label{def:lossy:min_rate}
  {\rm(Inf lossy CD-achievable rate)}
  \begin{eqnarray*}
    \lefteqn{\hspace{-3mm}
             R(X,Y|D_X,D_Y) = \inf\bigl\{R| R\mbox{ is a lossy}}\\
    && \hspace{-3mm}\mbox{CD-achievable rate of }(X,Y)
       \mbox{ for }(D_X,D_Y)\bigr\}.
  \end{eqnarray*}
\end{teigi}

\subsection{Statement of results}
\label{sec:full:state}

\begin{teiri}  \label{theorem:main:lossy}
  {\rm(Lossy coding theorem for CD code)}
  \begin{eqnarray*}
    \lefteqn{\hspace{-4mm}R(X,Y|D_X,D_Y)}\\
    &\hspace{-4mm}=& \hspace{-2mm}
        \min_{P_{U|XY}\in\cP_{CD}(\cU|P_{XY})}\bigl[\max\{I(X;U|Y),I(Y;U|X)\}\bigr],
  \end{eqnarray*}
  where the alphabet $\cU$ satisfies
  \begin{eqnarray*}
    |\cU| &\le& |\cX\times\cY|+2
  \end{eqnarray*}
  and $\cP_{CD}(\cU|P_{XY})\subseteq\cP(\cU|P_{XY})$ is a set of probability
  distributions such that there exist functions $\phi_{(1)}:\cU\times\cY\to\wh{\cX}$ and
  $\phi_{(2)}:\cU\times\cX\to\wh{\cY}$ that satisfy
  \begin{eqnarray*}
    D_X &\ge& E\left[\Delta_X(X,\phi_{(1)}(U,Y))\right],\\
    D_Y &\ge& E\left[\Delta_Y(Y,\phi_{(2)}(U,X))\right].
  \end{eqnarray*}
\end{teiri}

\medskip
Several important relationships between Theorem \ref{theorem:main:lossy} and previously
reported results are presented in the following.

\begin{hodai}  \label{lemma:lossless}
  {\rm (Compatibility with the result obtained for the lossless coding)}\\
  Suppose that $\wh{\cX}=\cX$, $\wh{\cY}=\cY$, $\Delta_X(x,\wh{x})=0$ if and only if
  $x=\wh{x}$ and $\Delta_Y(y,\wh{y})=0$ if and only if $y=\wh{y}$. In this case, the inf
  achievable rate $R(X,Y|D_X,D_Y)$ for $D_X=D_Y=0$ is reduced to the minimum achievable
  rate for the lossless coding.
  \begin{eqnarray*}
    R(X,Y)
    &\eqdef& R(X,Y|D_X=0,D_Y=0)\\
    &  =   & \max\{H(X|Y),H(Y|X)\},
  \end{eqnarray*}
  which coincides with the result reported by Willems et al.
  \cite{BroadcastSatelliteCoding}.
\end{hodai}

\begin{proof}
Note that if the conditions shown in Lemma \ref{lemma:lossless} satisfy we have
\begin{eqnarray*}
  \rho_n^{(X)}=0 &\Longleftrightarrow&
  \Pr\{X^n\neq\wh{\varphi}_n^{(1)}(A_n,Y^n))\}=0,\\
  \rho_n^{(Y)}=0 &\Longleftrightarrow&
  \Pr\{Y^n\neq\wh{\varphi}_n^{(2)}(A_n,X^n))\}=0.
\end{eqnarray*}
\end{proof}

\begin{hodai} \label{lemma:conditional}
  {\rm (Relationship to the conditional rate-distortion function)}
  \begin{eqnarray*}
    R(X,Y|D_X=d_1,D_Y) &=& R_C(Y|X,D_Y),\\
    R(X,Y|D_X,D_Y=d_2) &=& R_C(X|Y,D_X)
  \end{eqnarray*}
  if $d_1\ge\ov{\Delta}_X$ and $d_2\ge\ov{\Delta}_Y$, where $R_C(X|Y,D)$ denotes the
  {\it conditional rate-distortion function} \cite{RateDistortion:berger}, namely the
  minimum achievable rate when $X$ is encoded and reproduced both with the side
  information $Y$ to guarantee the distortion criterion $D$. 
\end{hodai}

\begin{proof}
It is sufficient to show that first equation. The condition $d_1\ge\ov{\Delta}_X$ implies
that one of the two messages does not have to be reproduced. Therefore, the encoder
$\varphi_n$ sends the codeword only to the decoder $\wh{\varphi}_n^{(2)}$, which means
that the coding rate characterized by the conditional rate-distortion function is
an achievable rate.
\begin{eqnarray*}
  R(X,Y|D_X=d_1,D_Y) &\le& R_C(Y|X,D_Y).
\end{eqnarray*}
On the other hand, we have
\begin{eqnarray*}
  \lefteqn{R(X,Y|D_X,D_Y)}\\
  &\ge& \max\{R_C(X|Y,D_X),R_C(Y|X,D_Y)\}\\
  &\ge& R_C(Y|X,D_Y).
\end{eqnarray*}
from the result of Theorem \ref{theorem:main:lossy}.
\end{proof}

\begin{hodai} \label{lemma:wz}
  {\rm (Relationships to the conditional rate-distortion function and Wyner-Ziv rate
   distortion function)}
  \begin{eqnarray*}
    \lefteqn{\max\{R_C(X|Y,D_X),R_C(Y|X,D_Y)\}}\\
    &\le& R(X,Y|D_X,D_Y)\\
    &\le& \max\{R_{WZ}(X|Y,D_X),R_{WZ}(Y|X,D_Y)\},
  \end{eqnarray*}
  where $R_{WZ}(X|Y,D_X)$ is the minimum achievable rate for the coding system called
  the {\it Wyner-Ziv coding system} \cite{WynerZiv}, where $X$ is encoded without any
  side information and reproduced with the side information $Y$. 
\end{hodai}

\begin{proof}
The left inequality was shown in the proof of Lemma \ref{lemma:conditional}. The
right inequality was shown by Kuzuoka et al. \cite{LinearUniversalComplementaryPaper}.
\end{proof}

Lemma \ref{lemma:wz} indicates that there may be some rate losses only for the lossy
coding. This property results from the auxiliary random variable $U$ included in the inf
achievable rate $R(X,Y|D_X,D_Y)$.

\section{Extension to multiple sources}
\label{sec:many}

Theorem \ref{theorem:main:lossy} considered only two correlated sources. However, the
theorem can be easily extended to any finite number of correlated sources.

Let $\X$ be a set of $N$ discrete memoryless sources
\begin{eqnarray*}
  \X &=& \{X^{(1)},X^{(2)},\cdots,X^{(N)}\},
\end{eqnarray*}
each of which $X^{(i)}$ takes a value in a finite set $\cX^{(i)}$ $(i\in\cI_N)$. For a
given subset $\cS\subseteq\cI_N$ of source indexes, the corresponding subsets of
sources, alphabets and its members are denoted by
\begin{eqnarray*}
  \X^{(\cS)}  &=& \{X^{(i)}| i\in\cS\},\\
  \cX^{(\cS)} &=& \prod_{i\in\cS}\cX^{(i)},\\
  \x^{(\cS)}  &=& \{x^{(i)}\in\cX^{(i)}| i\in\cS\}.
\end{eqnarray*}
Similarly, for a given subset $\cS\subseteq\cI_N$, the $n$-th Cartesian product of
$\cX^{(\cS)}$, its member and the corresponding random variable are written as
$\cX^{(\cS)n}$, $\x^{(\cS)n}$ and $\X^{(\cS)n}$, respectively. A substring of
$\x^{(\cS)n}$ is written as $\x_{i}^{(\cS)j}$ for $i\le j$. With $\cS=\cI_N$, we
denote $\X^{(\cS)n}=\X^n$. Also for a given subset $\cS\subseteq\cI_N$, its complement is
denoted by $\cS^c=\cI_N-\cS$.

Here, we introduce the definition and the coding theorem of the {\it generalized
complementary delivery code} which considers multiple correlated sources, multiple
encoders and multiple decoders.

\begin{teigi}  \label{def:manysources:code}
  {\rm(GCD (Generalized Complementary Delivery) code)}\\
  A set $(\varphi_n,\wh{\varphi}_n^{(1)},\cdots,\wh{\varphi}_n^{(M)})$ of single encoder
  and $M$ decoders is a GCD code
  \[
    \left(n,M_n,\{\rho_n^{(j,i)}\}_{j\in\cI_M,i\in\cS_j}\right)
  \]
  for the source $\X$ if and only if for any $j\in\cI_M$ and $i\in\cS_j\subseteq\cI_N$
  \begin{eqnarray*}
    \varphi_n            &:& \cX^{(\cI_N)n}\rightarrow\cI_{M_n}\\
    \wh{\varphi}_n^{(j)} &:& \cI_{M_n}\times\cX^{(\cS_j^c)n}\to\wh{\cX}^{(\cS_j)n},\\
    \rho_n^{(j,i)}       &=& E\left[\Delta_{X^{(i)}}^n(X^{(i)n},
                             \wh{\varphi}_n^{(j;i)}(A_n,\X^{(\cS_j^c)n}))\right],\\
    A_n &=& \varphi_n(\X^n),
  \end{eqnarray*}
  where $\wh{\varphi}_n^{(j;i)}$ is the output of $\wh{\varphi}_n^{(j)}$ that corresponds
  to the reproduction of $X^{(i)n}$.
\end{teigi}

\begin{teigi}  \label{def:manysources:rate}
  {\rm(Lossy GCD-achievable rate)}\\
  $R$ is a lossy GCD-achievable rate of the source $\X$ for a given set
  \begin{eqnarray*}
    \D &=& \{D_{j,i}\}_{j\in\cI_M,i\in\cS_j}
  \end{eqnarray*}
  of distortion criteria if and only if there exists a sequence
  \[
    \left\{
      \left(n,M_n,\{\rho_n^{(j,i)}\}_{j\in\cI_M,i\in\cS_j}\right)
    \right\}_{n=1}^{\infty}
  \]
  of GCD codes for the source $\X$ such that for any $j\in\cI_M$ and $i\in\cS_j$
  \begin{eqnarray*}
    && \limsup_{n\to\infty}\frac 1n\log M_n \le R,\\
    && \limsup_{n\to\infty}\rho_n^{(j,i)} \le D_{j,i}.
  \end{eqnarray*}
\end{teigi}

\begin{teigi}  \label{def:manysources:min_rate}
  {\rm(Inf lossy GCD-achievable rate)}
  \begin{eqnarray*}
    \lefteqn{\hspace{-3mm}
             R(\X|\D) = \inf\bigl\{R| R\mbox{ is a lossy}}\\
    && \hspace{-3mm}\mbox{GCD-achievable rate of }\X
       \mbox{ for }\D\bigr\}.
  \end{eqnarray*}
\end{teigi}

\begin{teiri}  \label{theorem:manysources:lossy}
  {\rm(Coding theorem of lossy GCD code)}
  \begin{eqnarray*}
    \lefteqn{R(\X|\D)}\\
    &=& \min_{P_{U|\X}\in\cP_{CD}(\cU|P_{\X})}
        \max_{j\in\cI_M}I\left(\left.\X^{(\cS_j)};U\right|\X^{(\cS_j^c)}\right),
  \end{eqnarray*}
  where the alphabet $\cU$ satisfies
  \begin{eqnarray*}
    |\cU|\le|\cX^{(\cI_N)}|+\sum_{j=1}^M |\cS_j|,
  \end{eqnarray*}
  and $\cP_{CD}(\cU|P_{\X})\subseteq\cP(\cU|P_{\X})$ is a set of probability
  distributions such that for any $j\in\cI_M$ and $i\in\cS_j$ there exists a function
  $\phi_{(j,i)}:\cU\times\cX^{(\cS_j^c)}\to\wh{\cX}^{(i)}$ that satisfy
  \begin{eqnarray*}
    D_{j,i}
    &\ge& E\left[\Delta_{X^{(i)}}\left(X^{(i)},
          \phi_{(j,i)}\left(U,\X^{(\cS_j^c)}\right)\right)\right].
  \end{eqnarray*}
\end{teiri}

\medskip

As a typical example, Theorem \ref{theorem:manysources:lossy} can be applied to the
coding problem formulated by Willems et al. \cite{BroadcastSatelliteCoding}. In this
coding system, the encoder sends three messages $\X=\{X,Y,Z\}$ to three
decoders, and each decoder has access to one of three messages to reproduce the two other
messages. Theorem \ref{theorem:manysources:lossy} indicates that the inf achievable rate
for this coding problem is obtained as
\begin{eqnarray*}
  \lefteqn{R(X,Y,Z|D_1,D_2,D_3)}\\
  &=& \min_{P_{U|XYZ}\in\cP_{CD}(\cU|P_{XYZ})}\\
  & & \max\{I(XY;U|Z),I(YZ;U|X),I(XZ;U|Y)\},
\end{eqnarray*}
where the alphabet $\cU$ satisfies $|\cU|\le|\cX\times\cY\times\cZ|+6$, and
$\cP_{CD}(\cU|P_{XYZ})\subseteq\cP(\cU|P_{XYZ})$ is a set of probability distributions
such that there exist functions 
\begin{align*}
  \phi_{(12)} : \cU\times\cX\to\wh{\cY},&\quad
  \phi_{(13)} : \cU\times\cX\to\wh{\cZ},\\
  \phi_{(21)} : \cU\times\cY\to\wh{\cX},&\quad
  \phi_{(23)} : \cU\times\cY\to\wh{\cZ},\\
  \phi_{(31)} : \cU\times\cZ\to\wh{\cX},&\quad
  \phi_{(32)} : \cU\times\cZ\to\wh{\cY}
\end{align*}
that satisfy
\begin{eqnarray*}
  D_{12} &\ge& E[\Delta_Y(Y,\phi_{(12)}(U,X))],\\
  D_{13} &\ge& E[\Delta_Z(Z,\phi_{(13)}(U,X))],\\
  D_{21} &\ge& E[\Delta_X(X,\phi_{(21)}(U,Y))],\\
  D_{23} &\ge& E[\Delta_Z(Z,\phi_{(23)}(U,Y))],\\
  D_{31} &\ge& E[\Delta_X(X,\phi_{(31)}(U,Z))],\\
  D_{32} &\ge& E[\Delta_Y(Y,\phi_{(32)}(U,Z))].
\end{eqnarray*}

\section{Proof of theorems}
\label{sec:proof}

\subsection{Theorem \ref{theorem:main:lossy}: converse part}
\label{sec:proof:converse}

\begin{proof}~\\
Let a sequence $\{(\varphi_n,\wh{\varphi}^{(1)}_n,\wh{\varphi}^{(2)}_n)\}_{n=1}^{\infty}$
of CD codes be given that satisfy the conditions of Definitions \ref{def:lossy:code} and
\ref{def:lossy:rate}. From Definition \ref{def:lossy:rate}, for any $\delta>0$ there
exists an integer $n_1=n_1(\delta)$ and then for all $n\ge n_1(\delta)$, we can obtain
\begin{eqnarray*}
  \frac 1n\log M_n &\le& R+\delta. \label{eq:proof:1}
\end{eqnarray*}
It should be remembered that $A_n=\varphi_n(X^n,Y^n)$. Then, we obtain
\begin{eqnarray*}
  \lefteqn{n(R+\delta)}\\
  &\ge& \log M_n\\
  &\ge& H(A_n)\\
  &\ge& H(A_n|Y^n)\\
  & = & I(X^n;A_n|Y^n)\quad(\because A_n=\varphi_n(X^n,Y^n))\\
  & = & H(X^n|Y^n)-H(X^n|A_nY^n)\\
  & = & \sum_{k=1}^n \{H(X_k|Y_k)-H(X_k|A_nX^{k-1}Y^n)\}\\
  & = & \sum_{k=1}^n I(X_k;A_nX^{k-1}Y^{k-1}Y_{k+1}^{n}|Y_k)\\
  &\ge& \sum_{k=1}^n I(X_k;A_nX^{k-1}Y^{k-1}|Y_k).
\end{eqnarray*}
Let us define random variables $U_k=A_nX^{k-1}Y^{k-1}$. With these definitions, we have
\begin{eqnarray*}
  n(R+\delta) &\ge& \sum_{k=1}^n I(X_k;U_k|Y_k).
\end{eqnarray*}
In a similar manner, we obtain
\begin{eqnarray*}
  n(R+\delta) &\ge& \sum_{k=1}^n I(Y_k;U_k|X_k).
\end{eqnarray*}
Here, let $J$ be a random variable that is independent of $(X,Y)$ and uniformly
distributed over the set $\cI_n$. We define a random variable $U=(J,U_J)$.
This implies that
\begin{eqnarray*}
  \lefteqn{R+\delta}\nonumber\\
   &\ge& \frac 1n\sum_{k=1}^n I(X_k;U_k|Y_k)\\
   & = & \frac 1n\sum_{k=1}^n\{H(X_k|Y_k)-H(X_k|U_kY_k)\}\\
   & = & \frac 1n\sum_{k=1}^n\{H(X_k|Y_k)-H(X_J|U_JY_J,J=k)\}\nonumber\\
   & = & H(X|Y)-H(X_J|JU_JY_J)\\
   & = & H(X|Y)-H(X|UY)\\
   & = & I(X;U|Y)
\end{eqnarray*}
and
\begin{eqnarray*}
  R+\delta &\ge& I(Y;U|X).
\end{eqnarray*}
Since $\delta>0$ is arbitrary, we obtain
\begin{eqnarray*}
  R &\ge& \max\{I(X;U|Y),I(Y;U|X)\}.
\end{eqnarray*}

We next show the existence of functions $\phi_{(1)}$ and $\phi_{(2)}$ that satisfy
the conditions of Theorem \ref{theorem:main:lossy}. From Definition
\ref{def:lossy:rate}, for any $\gamma>0$, there
exists an integer $n_2=n_2(\gamma)$, and for all $n\ge n_2(\gamma)$, we have
\begin{eqnarray*}
  D_X+\gamma
  &\ge& \frac 1n \sum_{k=1}^n E\left[\Delta_X(X_k,
        \wh{\varphi}_{n,k}^{(1)}(A_n,Y^n))\right]\\
  & = & \frac 1n \sum_{k=1}^n E\left[\Delta_X(X_k,\wh{X}_k)\right],\\
  D_Y+\gamma
  &\ge& \frac 1n \sum_{k=1}^n E\left[\Delta_Y(Y_k,
        \wh{\varphi}_{n,k}^{(2)}(A_n,X^n))\right]\\
  & = & \frac 1n \sum_{k=1}^n E\left[\Delta_Y(Y_k,\wh{Y}_k)\right],
\end{eqnarray*}
where $\wh{\varphi}_{n,k}^{(i)}$ ($i=1,2$, $k\in\cI_n$) is the output of
$\wh{\varphi}_{(i)}^n$ at time $k$, and
\begin{eqnarray*}
  \wh{X}_k &=& \wh{\varphi}_{n,k}^{(1)}(A_n,Y^n),\\
  \wh{Y}_k &=& \wh{\varphi}_{n,k}^{(2)}(A_n,X^n).
\end{eqnarray*}
We note that $U_kY_k$ contains $A_nY^k$,
and $U_kX_k$ contains $A_nX^k$, which implies that $Y_{k+1}^n$ (resp. $X_{k+1}^n$) is
further needed to generate $\wh{X}_k$ from $U_kY_k$ (resp. $\wh{Y}_k$ from $U_kX_k$).
Here, let us define the distribution $Q_{k_1,k_2}$ of $A_nX^{k_1}Y^{k_2}$, namely for any
$x^{k_1}\in\cX^{k_1}$, $y^{k_2}\in\cY^{k_2}$ and $a_n\in\cI_{M_n^{(1)}}$
\begin{eqnarray*}
  \lefteqn{Q_{k_1,k_2}(a_n,x^{k_1},y^{k_2})}\\
  &\eqdef& \Pr\{\varphi_n(X^n,Y^n)=a_n, X^{k_1}=x^{k_1}, Y^{k_2}=y^{k_2}\}\\
  & = & \sum_{\stackrel{(x_{k_1+1}^{n},y_{k_2+1}^{n})\in\cX^{n-k_1}\times\cY^{n-k_2}:}
        {\varphi_n(x^n,y^n)=a_n}}P_{X^nY^n}(x^n,y^n).
\end{eqnarray*}
Also, let $Q_k^{(1)}$ be the distribution of $X_k$ given $U_kY_k$, namely for any
$u_k=a_nx^{k-1}y^{k-1}$
\begin{eqnarray*}
  Q_k^{(1)}(x_k|u_k,y_k)
  &\eqdef& \frac{Q_{k,k}(a_n,x^k,y^k)}{Q_{k-1,k}(a_n,x^{k-1},y^k)},
\end{eqnarray*}
and $Q_k^{(2)}$ be the distribution of $Y_k$ given $U_kX_k$ defined similarly.
\begin{eqnarray*}
  Q_k^{(2)}(y_k|u_k,x_k)
  &\eqdef& \frac{Q_{k,k}(a_n,x^k,y^k)}{Q_{k,k-1}(a_n,x^k,y^{k-1})}.
\end{eqnarray*}
Further, let us define $\wt{Y}_{k+1}^n(U_k,Y_k)$ (resp. $\wt{X}_{k+1}^n(U_k,X_k)$) as a
random variable selected to minimize the average distortion between $X_k$ and $\wh{X}_k$
given $U_kY_k$ (resp. between $Y_k$ and $\wh{Y}_k$ given $U_kX_k$), namely
\begin{eqnarray*}
  \lefteqn{\wt{Y}_{k+1}^n(U_k,Y_k) \eqdef
           {\displaystyle\mathop{\arg\min}_{Y_{k+1}^n\in\cY^{n-k}}}}\nonumber\\
  && \sum_{X_k\in\cX}Q_k^{(1)}(X_k|U_k,Y_k)\Delta_X(X_k,\wh{X}_k),\\
  \lefteqn{\wt{X}_{k+1}^n(U_k,X_k) \eqdef
           {\displaystyle\mathop{\arg\min}_{X_{k+1}^n\in\cX^{n-k}}}}\nonumber\\
  && \sum_{Y_k\in\cY}Q_k^{(2)}(Y_k|U_k,X_k)\Delta_Y(Y_k,\wh{Y}_k).
\end{eqnarray*}
We choose the functions $\phi_{(1)}$ and $\phi_{(2)}$ as follows:
\begin{eqnarray*}
  \phi_{(1)k}(U_k,Y_k)
  &\eqdef& \wh{\varphi}_{n,k}^{(1)}(A_n,Y^k*\wt{Y}_{k+1}^n(U_k,Y_k)),\\
  \phi_{(2)k}(U_k,X_k)
  &\eqdef& \wh{\varphi}_{n,k}^{(2)}(A_n,X^k*\wt{X}_{k+1}^n(U_k,X_k)),\\
  \phi_{(1)}(U,Y) &\eqdef& \phi_{(1)J}(U_J,Y),\\
  \phi_{(2)}(U,X) &\eqdef& \phi_{(2)J}(U_J,X)\,
\end{eqnarray*}
where $*$ is an operator that represents string concatenation.
It is easy to see that
\begin{eqnarray*}
  \lefteqn{E\left[\Delta_X(X_k,\phi_{(1)k}(U_k,Y_k))\right]}\nonumber\\
  & = & E\left[\Delta_X(X_k,\wh{\varphi}_{n,k}^{(1)}(A_n,Y^k*\wt{Y}_{k+1}^n(U_k,Y_k)))
         \right]\\
  &\le& E\left[\Delta_X(X_k,\wh{\varphi}_{n,k}^{(1)}(A_n,Y^n))\right]\\
  & = & E\left[\Delta_X(X_k,\wh{X}_k)\right]\\
  \lefteqn{E\left[\Delta_Y(Y_k,\phi_{(2)k}(U_k,X_k))\right]}\nonumber\\
  &\le& E\left[\Delta_Y(Y_k,\wh{Y}_k)\right].
\end{eqnarray*}
This implies
\begin{eqnarray*}
  D_X+\gamma &\ge& \frac 1n\sum_{k=1}^{n}E\left[\Delta_X(X_k,\wh{X}_k)\right]\\
             &\ge& \frac 1n \sum_{k=1}^n
                   E\left[\Delta_X(X_k,\phi_{(1)k}(U_k,Y_k))\right]\\
             & = & \frac 1n \sum_{k=1}^n
                   E\left[\Delta_X(X,\phi_{(1)J}(U_J,Y))|J=k\right]\\
             & = & E\left[\Delta_X(X,\phi_{(1)}(U,Y))\right],\\
  D_Y+\gamma &\ge& E\left[\Delta_Y(Y,\phi_{(2)}(U,X))\right].
\end{eqnarray*}
Since $\gamma>0$ is arbitrary, we obtain
\begin{eqnarray*}
  D_X &\ge& E\left[\Delta_X(X,\phi_{(1)}(U,Y))\right],\\
  D_Y &\ge& E\left[\Delta_Y(Y,\phi_{(2)}(U,X))\right].
\end{eqnarray*}

It remains to establish that the bound on $|\cU|$ specified in Theorem
\ref{theorem:main:lossy} does not affect the determination of the inf achievable rate
$R(X,Y|D_X,D_Y)$. To do this, we introduce the support lemma
\cite[Lemma 3.3.4]{CsiszarKorner}. We can see that
\begin{eqnarray}
  \lefteqn{P_{XY}(x,y) = \sum_{u\in\cU}P_U(u)P_{XY|U}(x,y|u),}\nonumber\\
  \lefteqn{I(X;U|Y) = H(X|Y)-H(X|UY)}\nonumber\\
  & = & H(X|Y)-\sum_{u\in\cU}P_U(u)\nonumber\\
  &   & \sum_{(x,y)\in\cX\times\cY}\hspace{-5mm}
        P_{XY|U}(x,y|u)\log\frac{P_{Y|U}(y|u)}{P_{XY|U}(x,y|u)}\nonumber\\
  \lefteqn{I(Y;U|X) = H(Y|X)-H(Y|UX)}\nonumber\\
  & = & H(Y|X)-\sum_{u\in\cU}P_U(u)\nonumber\\
  &   & \sum_{(x,y)\in\cX\times\cY}P_{XY|U}(x,y|u)
        \log\frac{P_{X|U}(x|u)}{P_{XY|U}(x,y|u)}\nonumber\\
  \lefteqn{E[\Delta_X(X,\phi_{(1)}(U,Y))]}\nonumber\\
  & = & \sum_{u\in\cU}P_U(u)\hspace{-5mm}\sum_{(x,y)\in\cX\times\cY}\hspace{-5mm}
        P_{XY|U}(x,y|u)\Delta_X(x,\phi_{(1)}(u,y))\nonumber\\
  &\ge& \sum_{u\in\cU}P_U(u)\sum_{y\in\cY}\min_{\wh{x}\in\wh{\cX}}\nonumber\\
  &   & \hspace{15mm}\sum_{x\in\cX}P_{XY|U}(x,y|u)\Delta_X(x,\wh{x}),\label{eq:proof:2}\\
  \lefteqn{E[\Delta_Y(Y,\phi_{(2)}(U,X))]}\nonumber\\
  & = & \sum_{u\in\cU}P_U(u)\hspace{-5mm}\sum_{(x,y)\in\cX\times\cY}\hspace{-5mm}
        P_{XY|U}(x,y|u)\Delta_Y(y,\phi_{(2)}(u,x))\nonumber\\
  &\ge& \sum_{u\in\cU}P_U(u)\sum_{x\in\cX}\min_{\wh{y}\in\wh{\cY}}\nonumber\\
  &   & \hspace{15mm}\sum_{y\in\cY}P_{XY|U}(x,y|u)\Delta_Y(y,\wh{y}),\label{eq:proof:3}
\end{eqnarray}
where Eq.(\ref{eq:proof:2}) (resp. Eq.(\ref{eq:proof:3})) comes from the fact that for
given letters $(u,y)\in\cU\times\cY$ (resp. $(u,x)\in\cU\times\cX$) the output of the
function $\phi_{(1)}$ (resp. $\phi_{(2)}$) can be selected so as to minimize the average
distortion. We then define the following functions of a generic distribution
$Q\in\cP(\cX\times\cY)$:
\begin{eqnarray*}
  \lefteqn{q_1(Q,(x,y)) = Q(x,y),}\\
  \lefteqn{q_2(Q)=\max\{q_{2,1}(Q),q_{2,2}(Q)\},}\\
  \lefteqn{q_{2,1}(Q)}\\
  &=& H(X|Y)-\sum_{(x,y)\in\cX\times\cY}\hspace{-4mm}Q(x,y)
      \log\frac{\displaystyle\sum_{x'\in\cX}Q(x',y)}{Q(x,y)},\\
  \lefteqn{q_{2,2}(Q)}\\
  &=& H(Y|X)-\sum_{(x,y)\in\cX\times\cY}\hspace{-4mm}Q(x,y)
      \log\frac{\displaystyle\sum_{y'\in\cY}Q(x,y')}{Q(x,y)},\\
  \lefteqn{q_{3,1}(Q)
  = \sum_{y\in\cY}\min_{\wh{x}\in\wh{\cX}}\sum_{x\in\cX}Q(x,y)\Delta_X(x,\wh{x})},\\
  \lefteqn{q_{3,2}(Q)
  = \sum_{x\in\cX}\min_{\wh{y}\in\wh{\cY}}\sum_{y\in\cY}Q(x,y)\Delta_Y(y,\wh{y})}.
\end{eqnarray*}
Note that $|\cX\times\cY|-1$ functions are necessary to preserve the distribution
$Q(x,y)$ and $2$ functions to preserve the average distortion characterized by the
generic distribution $Q$. From the support lemma, we can find a generic distribution
$\alpha\in\cP(\wt{\cU})$ such that $\wt{\cU}\subseteq\cU$,
$|\wt{\cU}|\le|\cX\times\cY|+2$ and the following equations are simultaneously satisfied:
\begin{eqnarray}
  && \sum_{u\in\wt{\cU}}\alpha(u)q_1(P_{XY|U}(\cdot|u),(x,y)) = P_{XY}(x,y),
     \label{eq:proof:4}\\
  && \sum_{u\in\wt{\cU}}\alpha(u)q_2(P_{XY|U}(\cdot|u))\nonumber\\
  && \hspace{15mm} = \max\{I(X;U|Y),I(Y;U|X)\},\nonumber\\
  && \sum_{u\in\wt{\cU}}\alpha(u)q_{3,1}(P_{XY|U}(\cdot|u))\nonumber\\
  && = \sum_{u\in\wt{\cU}}\alpha(u)\sum_{y\in\cY}\min_{\wh{x}\in\wh{\cX}}
       \sum_{x\in\cX}P_{XY|U}(x,y|u)\Delta_X(x,\wh{x}), \nonumber\\
  && \sum_{u\in\wt{\cU}}\alpha(u)q_{3,2}(P_{XY|U}(\cdot|u))\nonumber\\
  && = \sum_{u\in\wt{\cU}}\alpha(u)\sum_{x\in\cX}\min_{\wh{y}\in\wh{\cY}}
       \sum_{y\in\cY}P_{XY|U}(x,y|u)\Delta_Y(y,\wh{y}). \nonumber
\end{eqnarray}
Here, let us define functions $\phi_{(1)}^*: \wt{\cU}\times\cY\to\wh{\cX}$ and
$\phi_{(2)}^*: \wt{\cU}\times\cX\to\wh{\cY}$ that satisfy
\begin{eqnarray*}
  \phi_{(1)}^*(u,y)
  &=& {\displaystyle\mathop{\arg\min}_{\wh{x}\in\wh{\cX}}}
      \sum_{x\in\cX}P_{XY|U}(x,y|u)\Delta_X(x,\wh{x}),\\
  \phi_{(2)}^*(u,x)
  &=& {\displaystyle\mathop{\arg\min}_{\wh{y}\in\wh{\cY}}}
      \sum_{y\in\cY}P_{XY|U}(x,y|u)\Delta_Y(y,\wh{y}).
\end{eqnarray*}
With these definitions, we have
\begin{eqnarray*}
  \hspace{-4mm}\sum_{u\in\wt{\cU}}\alpha(u)q_{3,1}(P_{XY|U}(\cdot|u))
  & = & E[\Delta_X(X,\phi_{(1)}^*(U,Y))],\\
  \hspace{-4mm}\sum_{u\in\wt{\cU}}\alpha(u)q_{3,2}(P_{XY|U}(\cdot|u))
  & = & E[\Delta_Y(Y,\phi_{(2)}^*(U,X))],
\end{eqnarray*}
and
\begin{eqnarray*}
  D_1
  &\ge& E[\Delta_X(X,\phi_{(1)}(U,Y))]\\
  &\ge& E[\Delta_X(X,\phi_{(1)}^*(U,Y))],\\
  D_2
  &\ge& E[\Delta_Y(Y,\phi_{(2)}(U,X))]\\
  &\ge& E[\Delta_Y(Y,\phi_{(2)}^*(U,X))].
\end{eqnarray*}
Hence, $\phi_{(1)}^*$ and $\phi_{(2)}^*$ satisfy the conditions of Theorem
\ref{theorem:main:lossy}. Further, Eq.(\ref{eq:proof:4}) implies that there exist a
random variable $\wt{U}$ and a joint distribution $P_{\wt{U}XY}$ that satisfy
\begin{eqnarray*}
  \alpha(u)P_{XY|U}(x,y|u) &=& P_{\wt{U}XY}(u,x,y)
\end{eqnarray*}
for all $(u,x,y)\in\wt{\cU}\times\cX\times\cY$. The new joint
distribution preserves the distribution $P_{XY}$
\begin{eqnarray*}
  \sum_{u\in\wt{\cU}}P_{\wt{U}XY}(u,x,y)
  &=& \sum_{u\in\wt{\cU}}\alpha(u)P_{XY|U}(x,y|u)\\
  &=& P_{XY}(x,y).
\end{eqnarray*}

This completes the proof of the converse part.
\end{proof}

\subsection{Theorem \ref{theorem:main:lossy}: direct part}
\label{sec:proof:achieve}

We begin by establishing some notation and mentioning a few basic facts that will be
used hereafter.

\begin{teigi}  \label{def:typical}
  {\rm(Set of typical sequences)}\\
  For any $\delta>0$, define the set of typical sequences as
  \begin{eqnarray*}
    \lefteqn{T_X^n(\delta) =}\nonumber\\
    &&  \Bigl\{x^n\in\cX^n: \left.\left|\frac{1}{n}N(x|x^n)-P_X(x)\right|\le\delta
        ~\forall x\in\cX\right\},
  \end{eqnarray*}
  where $N(x|x^n)$ stands for the number of occurrences of the letter $x$ included in the
  sequence $x^n$. A similar convention is used for other random variables. When the
  dimension is clear from the context, the superscript $n$ will be omitted, e.g.
  $T_X(\delta)$.
\end{teigi}

\begin{hodai}  \label{lemma:typical_prob}
  {\rm(Csisz\dash{a}r-K\dotdot{o}rner \cite{CsiszarKorner})}\\
  For any $\delta>0$
  \begin{eqnarray*}
    \Pr\{X^n\in T_X(\delta)\}      &\ge& 1-\epsilon_n(\delta),
  \end{eqnarray*}
  where
  \begin{eqnarray*}
    \lim_{n\to\infty}\epsilon_n(\delta) &=& 0.
  \end{eqnarray*}
\end{hodai}

\begin{hodai}  \label{lemma:typical_extend}
  {\rm(Csisz\dash{a}r-K\dotdot{o}rner \cite[Lemma 1.2.10]{CsiszarKorner})}\\
  For any $\delta,\delta'>0$, if $(x^n,y^n)\in T_{XY}(\delta_1)$ then
  $x^n\in T_X(\delta_1|\cY|)$.
\end{hodai}

\begin{hodai}  \label{lemma:csiszar1}
  {\rm(Steinberg-Merhav \cite{SuccessiveRefinementWithSI})}\\
  For any $\delta'>\delta>0$ and $x^n\in T_X(\delta)$
  \begin{eqnarray*}
    \lefteqn{\exp\{-n(I(X;U)+\epsilon_1)\}}\\
    && \le\sum_{u^n:(u^n,x^n)\in T_{UX}(\delta')}\hspace{-6mm}P_U(u^n)
       \le\exp\{-n(I(X;U)-\epsilon_2)\},
    \label{eq:csiszar1}
  \end{eqnarray*}
  where $\epsilon_1$ is a function of $(\delta,\delta')$, $\epsilon_2$ is a function of
  $(\delta,\delta')$ and
  \begin{eqnarray*}
    \lim_{\delta,\delta'\to 0}\epsilon_1 &=&
    \lim_{\delta,\delta'\to 0}\epsilon_2 = 0.
  \end{eqnarray*}
\end{hodai}

Now, we proceed with the proof of the direct part of Theorem \ref{theorem:main:lossy}.

\begin{proof}~\\
Let a distortion pair $(D_X,D_Y)$ be given, and
$P_{U|XY}\in\cP_{CD}(\cU|P_{XY})$. Fix arbitrary $\gamma,\delta>0$.

\medskip\noindent
\underline{\it Codeword selection: $\varphi_n$}\\
(1) Randomly generate $M_U$ independent codewords $u^n(i)\in\cU^n$ $(i\in\cI_{M_U})$, each
    of length $n$, according to $P_U$ to create a codebook
    $\cA_U=\{u^n(i)\}_{i=1}^{M_U}$.\\
(2) Partition the codebook $\cA_U$ into $N_U$ bins, each containing $L_U=M_U/N_U$ members
    of $\cA_U$. For simplicity, $M_U$ is a multiple of $N_U$. Let $\cA_U(j)$ denote the
    subset of $\cA_U$ whose elements are assigned to bin $j$
    ($j\in\cI_{N_U}$). Without loss of generality, we define
    \begin{eqnarray*}
      \cA_U(j) &=& \{u^n(i)\}_{i=(j-1)L_U+1}^{jL_U}.
    \end{eqnarray*}

\noindent
\underline{\it Encoding: $\varphi_n$}\\
(1) For a given input pair $(x^n,y^n)\in\cX^n\times\cY^n$ of sequences, the encoder seeks
    a vector $u^n\in\cA_U$ that satisfies $(u^n,x^n,y^n)\in T_{UXY}(k_1\delta)$, where
    $k_1>0$. If there is more than one such vector in the codebook $\cA_U$, the first one
    is chosen. If there is no such vector in the codebook $\cA_U$, a default vector is
    chosen, say $u^n(1)$, and an error is declared. The selected vector is denoted by
    $u^n(x^n,y^n)$.\\
(2) The value assigned to the encoder $\varphi_n(\cdot)$ is the bin index to which
    $u^n(x^n,y^n)$ belongs, that is,
    \begin{eqnarray*}
      \varphi_n(x^n,y^n) &=& j \quad\mbox{if } u^n(x^n,y^n)\in\cA_U(j).
    \end{eqnarray*}

\noindent
\underline{\it Decoding: $\wh{\varphi}^{(1)}_n$}\\
(1) The decoder has access to the bin index $j_U\in\cI_{N_U}$ received from the encoder
    and the sequence $y^n\in\cY^n$ of side information.\\
(2) The decoder seeks a unique vector $u^n\in\cA_U(j_U)$ that satisfies
    $(u^n,y^n)\in T_{UY}(k_2\delta)$, where $k_2>0$. This vector is denoted by
    $\wh{u}^n(y^n)$. If there is no or more than one
    vector $u^n\in\cA_U(j_U)$ jointly typical with $y^n$, arbitrary $\wh{u}^n$ is
    chosen, and an error is declared.\\
(3) The reconstruction vector $\wh{x}^n = (\wh{x}_1,\wh{x}_2,\cdots,\wh{x}_n)$
    is given by
    \begin{eqnarray*}
      \wh{x}_k &=& \phi_{(1)}(\wh{u}_k(y^n),\wh{y}_k)\quad(k\in\cI_n),
    \end{eqnarray*}
    where $\wh{u}_k(y^n)$ is the $k$-th element of $u^n(y^n)$.

\noindent
\underline{\it Decoding: $\wh{\varphi}^{(2)}_n$}\\
(1) The decoder has access to the bin index $j_U\in\cI_{N_U}$ and
    the sequence $x^n\in\cX^n$ of side information.\\
(2) In a similar manner to $\wh{\varphi}^{(1)}_n$, the decoder seeks a unique
    vector $u^n\in\cA_U(j_U)$ that satisfies $(u^n,x^n)\in T_{UX}(k_3\delta)$,
    where $k_3>0$, and the reconstruction vector $\wh{y}^n$ is given by
    \begin{eqnarray*}
      \wh{y}_k &=& \phi_{(2)}(\wh{u}_k(x^n),\wh{x}_k)\quad(k\in\cI_n).
    \end{eqnarray*}

\noindent
\underline{\it Distortion evaluation: $\wh{\varphi}^{(1)}_n$}\\
  For the distortion, we obtain
  \begin{eqnarray*}
    \lefteqn{\Delta_X^n(x^n,\wh{x}^n)}\nonumber\\
    &=& \frac 1n\sum_{k=1}^n\Delta_X(x_k,\wh{x}_k)\\
    &=& \frac 1n\sum_{k=1}^n\Delta_X(x_k,\phi_{(1)}(\wh{u}_k(y^n),\wh{y}_k))\\
    &=& \frac 1n\hspace{-12mm}\sum_{\hspace{12mm}(u,x,y)\in\cU\times\cX\times\cY}
        \hspace{-13mm}N(u,x,y|\wh{u}^n(y^n),x^n,y^n)\\
    & & \hspace{15mm}\Delta_X(x,\phi_{(1)}(u,y)).
  \end{eqnarray*}
  We note that $(u^n(x^n,y^n),x^n,y^n)\in T_{UXY}(k_1\delta)$. Also, if no error occurs in
  the encoding/decoding process, we have $u^n(x^n,y^n)=\wh{u}^n(y^n)$. In this case,
  the following inequalities are satisfied:
  \begin{eqnarray*}
    \lefteqn{\Delta_X^n(x^n,\wh{x}^n)}\nonumber\\
    &\le& \hspace{-12mm}\sum_{\hspace{12mm}(u,x,y)\in\cU\times\cX\times\cY}
          \hspace{-12mm}(P_{UXY}(u,x,y)+k_1\delta)\Delta_X(x,\phi_{(1)}(u,y))\\
    &\le& E\left[\Delta_X(X,\phi_{(1)}(U,Y))\right]
          +k_1\delta\overline{\Delta}_X|\cU\times\cX\times\cY|\\
    &\le& D_X+k_1\delta\overline{\Delta}_X|\cU\times\cX\times\cY|.
  \end{eqnarray*}
  We denote error probabilities in the encoding/decoding process as $P_e^n$. Then,
  the average distortion can be bounded as
  \begin{eqnarray*}
    \lefteqn{\hspace{-3mm}E\left[\Delta_X^n(X^n,\wh{X}^n)\right]}\\
    &\hspace{-8mm}\le&
      \hspace{-3mm}
      (1-P_e^n)(D_X+k_1\delta\overline{\Delta}_X|\cU\times\cX\times\cY|)+
      P_e^n\overline{\Delta}_X.
  \end{eqnarray*}
  Since $\delta>0$ is arbitrarily small for a sufficiently large $n$, if $P_e^n$ vanishes
  as $n\to\infty$, we can obtain
  \begin{eqnarray*}
    \limsup_{n\to\infty}E\left[\Delta_X^n(X^n,\wh{X}^n)\right] &\le& D_X.
  \end{eqnarray*}
  
\noindent
\underline{\it Distortion evaluation: $\wh{\varphi}^{(2)}_n$}\\
  We can obtain
  \begin{eqnarray*}
    \limsup_{n\to\infty}E\left[\Delta_Y^n(Y^n,\wh{Y}^n)\right] &\le& D_Y
  \end{eqnarray*}
  in a similar manner to $\wh{\varphi}^{(1)}_n$.

\medskip\noindent
\underline{\it Error evaluation: $\varphi_n$}\\
  If there is no $u^n\in\cA_U$ that satisfies $(u^n,x^n,y^n)\in T_{UXY}(k_1\delta)$,
  an encoding error has occurred. This event is denoted as
  \begin{eqnarray*}
    E_{11} &\eqdef&
      \bigcap_{i=1}^{M_U}\left\{(u^n(i),x^n,y^n)\notin T_{UXY}(k_1\delta)\right\}.
  \end{eqnarray*}
  Here, let us define
  \begin{eqnarray*}
    E_0 &\eqdef& \{(x^n,y^n)\in T_{XY}(k_0\delta)\},
  \end{eqnarray*}
  where $k_0>0$. From Lemma \ref{lemma:typical_prob}, $\Pr\{E_0^c\}\to 0$ as
  $n\to\infty$. Then, we have
  \begin{eqnarray*}
    \Pr\{E_{1}\}
    &\le& \Pr\{E_{1}\cup E_{0}^c\}\\
    & = & \Pr\{E_0^c\}+\Pr\{E_{0}\cap E_{1}\},
  \end{eqnarray*}
  \begin{eqnarray*}
    \lefteqn{\Pr\{E_0\cap E_1\}}\nonumber\\
    &\le& \sum_{(x^n,y^n)\in T_{XY}(k_0\delta)}\hspace{-6mm}P_{XY}(x^n,y^n)\\
    &   & \Pr\left\{\left.\bigcap_{i=1}^{M_U}
          \{(U^n(i),x^n,y^n)\notin T_{UXY}(k_1\delta)\}\right|x^n,y^n\right\}\\
    & = & \sum_{(x^n,y^n)\in T_{XY}(k_0\delta)}\hspace{-6mm}P_{XY}(x^n,y^n)\\
    &   & \Pr\left\{\bigcap_{i=1}^{M_U}
          \left\{(U^n(i),x^n,y^n)\notin T_{UXY}(k_1\delta)\right\}\right\}\\
    &   & \quad(\because u^n(i)\mbox{ is selected independently of $(x^n,y^n)$})
          \nonumber\\
    &\le& \sum_{(x^n,y^n)\in T_{XY}(k_0\delta)}\hspace{-6mm}P_{XY}(x^n,y^n)\\
    &   & \left[1-\exp\left\{-n(I(XY;U)+\epsilon_u)\right\}\right]^{M_U}\\
    &   & \quad(\because \mbox{Lemma \ref{lemma:csiszar1}})\nonumber\\
    &\le& \sum_{(x^n,y^n)\in T_{XY}(k_0\delta)}\hspace{-6mm}P_{XY}(x^n,y^n)\\
    &   & \exp\left[-M_U\exp\left\{-n(I(XY;U)+\epsilon_u)\right\}\right],\nonumber\\
    &   & \quad(\because (1-a)^n\le\exp(-an))
  \end{eqnarray*}
  where $\epsilon_u$ is a function of $(k_1\delta,k_0\delta)$. By setting $M_U$, $k_1$
  and $k_0$ as
  \begin{eqnarray*}
    M_U\ge\exp\{n(I(XY;U)+m_1\gamma)\},~m_1>0,
  \end{eqnarray*}
  $m_1\gamma>\epsilon_u$ and $k_1<k_0$, we have $\lim_{n\to\infty}\Pr\{E_{1}\}=0$.

\medskip\noindent
\underline{\it Error evaluation: $\wh{\varphi}^{(1)}_n$}\\
  If there is no or more than one $u^n\in\cA_U(j_U)$ such that
  $(u^n,y^n)\in T_{UY}(k_2\delta)$, a decoding error is declared.
  This event is classified into two cases.\\
  (1) First case: $(u^n(x^n,y^n),y^n)\notin T_{UY}(k_2\delta)$. However, this error
      does not occur by setting $k_2$ as $k_2>k_1|\cX|$ because $(u^n(x^n,y^n),x^n,y^n)$
      $\in T_{UXY}(k_1\delta)$ and Lemma \ref{lemma:typical_extend}.\\
  (2) Second case: If there exists $u^n\in\cA_U(j_U)$, $u^n\neq u^n(x^n,y^n)$ such that
      $(u^n,y^n)\in T_{UY}(k_2\delta)$. This event is denoted as
      \begin{eqnarray*}
        E_2 &\eqdef&
            \bigcup_{u^n\in\cA_U(j_U),u^n\neq u^n(x^n,y^n)}
	    \hspace{-7mm}\{(u^n,y^n)\in T_{UY}(k_2\delta)\}.
      \end{eqnarray*}
      Let $i(j,k)$ be the index $i$ of $k$-th $u^n(i)$ in $\cA_U(j)$, namely from the
      definition of $\cA_U(j)$ we have
      \begin{eqnarray*}
        i(j,k) &=& (j-1)L_U+k.
      \end{eqnarray*}
      Since if $(x^n,y^n)\in T_{XY}(k_0\delta)$ then $y^n\in T_Y(k_0\delta|\cX|)$, we
      have
      \begin{eqnarray*}
	\Pr\{E_{2}\}
	&\le& \Pr\{E_{2}\cup E_{0}^c\}\\
	& = & \Pr\{E_0^c\}+\Pr\{E_{0}\cap E_{2}\},
      \end{eqnarray*}
      \begin{eqnarray*}
	\lefteqn{\Pr\{E_0\cap E_2\}}\nonumber\\
        &\le& \sum_{k=1}^{L_U}\hspace{-2mm}
	      \sum_{y^n\in T_Y(k_0\delta|\cX|)}\hspace{-4mm}P_Y(y^n)\\
        &   & \hspace{10mm}\Pr\left\{(U^n(i(j_U,k)),y^n)\in T_{UY}(k_2\delta)\right\}\\
        &   & \quad(\because u^n(i) \mbox{ is selected independently of }y^n)\nonumber\\
        &\le& L_U\exp\{-n(I(Y;U)-\epsilon)\},\\
        &   & \quad(\because \mbox{Lemma }\ref{lemma:csiszar1})
     \end{eqnarray*}
     where $\epsilon$ is a function of $(k_0\delta\|\cX\|,k_2\delta)$. By setting $L_U$
     and $k_2$ as
     \begin{eqnarray*}
       L_U\le\exp\{n(I(Y;U)-l_1\gamma)\},~l_1>0,
     \end{eqnarray*}
     $l_1\gamma>\epsilon$ and $k_0|\cX|<k_2$, we have $\lim_{n\to\infty}\Pr\{E_2\}=0$.

\medskip\noindent
\underline{\it Error evaluation: $\wh{\varphi}^{(2)}_n$}\\
  This is almost the same as the case of $\wh{\varphi}^{(1)}_n$. We have to set
  \begin{eqnarray*}
    L_U &\le& \exp\{n(I(X;U)   -l_{2}\gamma)\},~l_{2}>0
  \end{eqnarray*}
  to vanish the encoding/decoding errors.

\medskip\noindent
\underline{\it Rate evaluation: $\varphi_n$}\\
  The encoder sends the indexes of the bin using
  \begin{eqnarray*}
    \lefteqn{\hspace{-3mm}R = \frac 1n\log N_U}\nonumber\\
    &\hspace{-3mm} = & \frac 1n\log\frac{M_U}{L_U}\\
    &\hspace{-3mm}\ge& I(XY;U)+m_{1}\gamma\nonumber\\
    &\hspace{-3mm}   & -\min\{I(Y;U)-l_{1}\gamma,I(X;U)-l_{2}\gamma\}\\
    &\hspace{-3mm} = & \max\{I(X;U|Y)+l_{1}\gamma,I(Y;U|X)+l_{2}\gamma\}+m_{1}\gamma
  \end{eqnarray*}
  bits per letter. Since $\gamma>0$ is arbitrary, we obtain the coding rate as
  $\max\{I(X;U|Y),I(Y;U|X)\}$.

This completes the proof of Theorem \ref{theorem:main:lossy}.
\end{proof}

\subsection{Theorem \ref{theorem:manysources:lossy}: converse part}
\label{sec:proof:multi:converse}

\begin{proof}~\\
The proof of Theorem \ref{theorem:manysources:lossy} is quite similar to that of Theorem
\ref{theorem:main:lossy}. Let a sequence
$\{(\varphi_n,\wh{\varphi}_n^{(1)},\cdots,\wh{\varphi}_n^{(M)})\}_{n=1}^{\infty}$
of GCD codes be given that satisfy the conditions of Definitions
\ref{def:manysources:code} and \ref{def:manysources:rate}. From Definition
\ref{def:manysources:rate}, for any $\delta>0$ there exists an integer $n_1=n_1(\delta)$
such that for all $n\ge n_1(\delta)$, we can obtain
\begin{eqnarray*}
  \frac 1n\log M_n &\le& R+\delta.
\end{eqnarray*}
In a similar manner to Theorem \ref{theorem:main:lossy} we obtain
\begin{eqnarray*}
  n(R+\delta) &\ge& \sum_{k=1}^n I(\X^{(\cS_j^c)}_k;A_n\X^{k-1}|\X^{(\cS_j^c)}_k).
\end{eqnarray*}
Let us define random variables $U_k=A_n\X^{k-1}$, and let $J$ be a random variable that
is independent of $\X$ and uniformly distributed over the set $\cI_n$. We define a random
variable $U=(J,U_J)$. This implies that for every $j\in\cI_M$
\begin{eqnarray*}
  R+\delta &\ge& I(\X^{(\cS_j)};U|\X^{(\cS_j^c)}).
\end{eqnarray*}
Since $\delta>0$ is arbitrary for a sufficiently large $n$, we obtain
\begin{eqnarray*}
  R &\ge& \max_{j\in\cI_M}I(\X^{(\cS_j)};U|\X^{(\cS_j^c)}).
\end{eqnarray*}

We next show the existence of functions $\phi_{(j,i)}$ $(j\in\cI_M$, $i\in\cS_j)$ that
satisfy the conditions of Theorem \ref{theorem:manysources:lossy}. From Definition
\ref{def:manysources:rate}, for any $\gamma>0$ there exists an integer $n_2=n_2(\gamma)$
such that for all $n\ge n_2(\gamma)$
\begin{eqnarray*}
  \lefteqn{D_{j,i}+\gamma}\\
  &\ge& \frac 1n \sum_{k=1}^n E\left[\Delta_{X^{(i)}}(X^{(i)}_k,
        \wh{\varphi}_{n,k}^{(j;i)}(A_n,\X^{(\cS_j^c)n}))\right],
\end{eqnarray*}
where $\wh{\varphi}_{n,k}^{(j;i)}$ ($k\in\cI_n$) is the output of
$\wh{\varphi}_n^{(j;i)}$ at time $k$. We note that $U_k\X^{(\cS_j^c)}_k$ contains
$A_n\X^{(\cS_j^c)k}$, which implies that $\X^{(\cS_j^c)n}_{k+1}$ is further needed to
generate $\wh{\X}^{(\cS_j)}_k$ from $U_k\X^{(\cS_j^c)}_k$. Here, let us define the
distribution $Q_{k_1,k_2}$ of $A_n\X^{(\cS_j)k_1}\X^{(\cS_j^c)k_2}$, namely for any
$\x^{(\cS_j)k_1}\in\cX^{(\cS_j)k_1}$, $\x^{(\cS_j^c)k_2}\in\cX^{(\cS_j^c)k_2}$ and
$a_n\in\cI_{M_n}$
\begin{eqnarray*}
  \lefteqn{Q_{k_1,k_2}(a_n,\x^{(\cS_j)k_1},\x^{(\cS_j^c)k_2})}\\
  &\eqdef& \Pr\{\varphi_n(\X^n)=a_n,\\
  &   & \hspace{7mm}
        \X^{(\cS_j)k_1}=\x^{(\cS_j)k_1},\X^{(\cS_j^c)k_2}=\x^{(\cS_j^c)k_2}\}\\
  & = & \sum_{\stackrel{\left(\x^{(\cS_j)n}_{k_1+1},\x^{(\cS_j^c)n}_{k_2+1}\right)\in
	\cX^{(\cS_j)n-k_1}\times\cX^{(\cS_j^c)n-k_2}:}
        {\varphi_n(\x^{(\cI_N)n})=a_n}}\hspace{-15mm}P_{\X^n}(\x^{(\cI_N)n}).
\end{eqnarray*}
Also, let $Q_k^{(j)}$ be the distribution of $\X^{(\cS_j)}_k$ given
$U_k\X^{(\cS_j^c)}_k$, namely for any $u_k=a_n\x^{(\cI_N)k-1}$
\begin{eqnarray*}
  \lefteqn{Q_k^{(j)}(\x^{(\cS_j)}_k|u_k,\x^{(\cS_j^c)}_k)}\\
  &\eqdef& \frac{Q_{k,k}  (a_n,\x^{(\cS_j)k},  \x^{(\cS_j^c)k})}
                {Q_{k-1,k}(a_n,\x^{(\cS_j)k-1},\x^{(\cS_j^c)k})}.
\end{eqnarray*}
Further, let us define $\wt{\X}^{(\cS_j^c)n}_{k+1}(U_k,\X^{(\cS_j^c)}_k,i)$ as random
variables selected to minimize the average distortion between $X^{(i)}_k$ and the output
of $\wh{\varphi}_{n,k}^{(j;i)}$ $(i\in\cS_j)$ given $U_k\X^{(\cS_j^c)}_k$, namely
\begin{eqnarray*}
  \lefteqn{\wt{\X}^{(\cS_j^c)n}_{k+1}(U_k,\X^{(\cS_j^c)}_k,i) \eqdef
           {\displaystyle\mathop{\arg\min}_{\X^{(\cS_j^c)n}_{k+1}\in
	   \cX^{(\cS_j^c)n-k}}}}\nonumber\\
  && \sum_{\X^{(\cS_j)}_k\in\cX^{(\cS_j)}}Q_k^{(j)}(\X^{(\cS_j)}_k|U_k\X^{(\cS_j^c)}_k)\\
  && \hspace{13mm}\Delta_{X^{(i)}}(\X^{(i)}_k,
     \wh{\varphi}_{n,k}^{(j;i)}(A_n,\X^{(\cS_j^c)n})).
\end{eqnarray*}
We choose the functions $\phi_{(j;i)}$ as follows:
\begin{eqnarray*}
  \lefteqn{\phi_{(j;i)k}(U_k,\X^{(\cS_j^c)}_k)}\\
  &\eqdef& \wh{\varphi}_{n,k}^{(j;i)}(A_n,\X^{(\cS_j^c)k}*\wt{\X}^{(\cS_j^c)n}_{k+1}
           (U_k,\X^{(\cS_j^c)}_k,i)),\\
  \lefteqn{\phi_{(j;i)}(U,\X^{(\cS_j^c)}) \eqdef \phi_{(j;i)J}(U_J,\X^{(\cS_j^c)})}
\end{eqnarray*}
In a similar way to Theorem \ref{theorem:main:lossy}, we obtain
\begin{eqnarray*}
  \lefteqn{D_{j,i}+\gamma}\\
  &\ge& \frac 1n \sum_{k=1}^{n}E\left[\Delta_{X^{(i)}}(X^{(i)}_k,
        \wh{\varphi}_{n,k}^{(j;i)}(A_n,\X^{(\cS_j^c)n}))\right]\\
  &\ge& \frac 1n \sum_{k=1}^n
        E\left[\Delta_{X^{(i)}}(X^{(i)}_k,\phi_{(j;i)k}(U_k,\X^{(\cS_j^c)}_k))\right]\\
  & = & E\left[\Delta_{X^{(i)}}(X^{(i)},\phi_{(j;i)}(U,\X^{(\cS_j^c)}))\right].
\end{eqnarray*}
Since $\gamma>0$ is arbitrary for a sufficiently large $n$, we obtain
\begin{eqnarray*}
  D_{j,i} &\ge& E\left[\Delta_{X^{(i)}}(X^{(i)},\phi_{(j;i)}(U,\X^{(\cS_j^c)}))\right].
\end{eqnarray*}

It remains to establish that the bound on $|\cU|$ specified in Theorem
\ref{theorem:manysources:lossy} does not affect the determination of the inf achievable
rate $R(\X|\D)$. In a similar way to Theorem \ref{theorem:main:lossy}, we then define the
following functions of a generic distribution $Q\in\cP(\cX^{(\cI_N)})$:
\begin{eqnarray*}
  \lefteqn{q_1(Q,\x^{(\cI_N)}) = Q(\x^{(\cI_N)})}\\
  \lefteqn{q_{2}(Q)=\max_{j\in\cI_M}q_{2,j}(Q),}\\
  \lefteqn{q_{2,j}(Q) = H(\X^{(\cS_j)}|\X^{(\cS_j^c)})}\\
  & & -\hspace{-4mm}\sum_{\x^{(\cI_N)}\in\cX^{(\cI_N)}}\hspace{-4mm}
      Q(\x^{(\cI_N)})\log\frac{\displaystyle
      \sum_{\wt{\x}^{(\cS_j)}\in\cX^{(\cS_j)}}\hspace{-4mm}
      Q(\wt{\x}^{(\cS_j)},\x^{(\cS_j^c)})}{Q(\x^{(\cI_N)})},\\
  \lefteqn{q_{3,m(j,i)}(Q)
  = \sum_{\x^{(\cS_j^c)}\in\cX^{(\cS_j^c)}}\min_{\wh{x}^{(i)}\in\wh{\cX}^{(i)}}}\\
  & & \sum_{\x^{(\cS_j)}\in\cX^{(\cS_j)}}
      Q(\x^{(\cI_N)})\Delta_{X^{(i)}}(x^{(i)},\wh{x}^{(i)}),
\end{eqnarray*}
where $j\in\cI_M$ $i\in\cS_j$ and $m(j,i)$ denotes the serial number of the source $X_i$
contained in the index set $\cS_j$ defined as follows:
\begin{eqnarray*}
  m(j,i) &\eqdef&
  \left|\{\wt{i}\in\cS_j| \wt{i}\le i\}\right|+\sum_{\wt{j}=1}^{j-1}|\cS_j|.
\end{eqnarray*}
Note that $|\cX^{(\cI_N)}|-1$ functions are needed to preserve the distribution
$Q(\x^{(\cI_N)})$, and $\sum_{j=1}^M|\cS_j|$ functions to preserve the average distortion
characterized by the generic distribution $Q$. From the support lemma, we can find a
generic distribution $\alpha\in\cP(\wt{\cU})$ such that $\wt{\cU}\subseteq\cU$,
\begin{eqnarray*}
  |\wt{\cU}| &\le& |\cX^{(\cI_N)}|+\sum_{j=1}^M|\cS_j|
\end{eqnarray*}
and the following equations are simultaneously satisfied:
\begin{eqnarray}
  && \sum_{u\in\wt{\cU}}\alpha(u)q_1(P_{\X|U}(\cdot|u),\x^{(\cI_N)})
     = P_{\X}(\x^{(\cI_N)}), \label{eq:proof:multi4}\\
  && \sum_{u\in\wt{\cU}}\alpha(u)q_2(P_{\X|U}(\cdot|u))\nonumber\\
  && \hspace{15mm} = \max_{j\in\cI_M}I(\X^{(\cS_j)};U|\X^{(\cS_j^c)}),\nonumber\\
  && \sum_{u\in\wt{\cU}}\alpha(u)q_{3,m(j,i)}(P_{\X|U}(\cdot|u))\nonumber\\
  && = \sum_{u\in\wt{\cU}}\alpha(u)\sum_{\x^{(\cS_j^c)}\in\cX^{(\cS_j^c)}}
       \min_{\wh{x}^{(i)}\in\wh{\cX}^{(i)}}\nonumber\\
  &&   \sum_{\x^{(\cS_j)}\in\cX^{(\cS_j)}}P_{\X|U}(\x^{(\cI_N)}|u)
       \Delta_{X^{(i)}}(x^{(i)},\wh{x}^{(i)}). \nonumber
\end{eqnarray}
Here, let us define functions $\phi_{(j;i)}^*:$
$\wt{\cU}\times\cX^{(\cS_j^c)}\to\wh{\cX}^{(i)}$ $(j\in\cI_M, i\in\cS_j)$ that satisfy
\begin{eqnarray*}
  \lefteqn{\phi_{(j;i)}^*(u,\x^{(\cS_j^c)})
    = {\displaystyle\mathop{\arg\min}_{\wh{x}^{(i)}\in\wh{\cX}^{(i)}}}}\\
  & & \sum_{\x^{(\cS_j)}\in\cX^{(\cS_j)}}P_{\X|U}(\x^{(\cI_N)}|u)
      \Delta_{X^{(i)}}(x^{(i)},\wh{x}^{(i)}).
\end{eqnarray*}
With these definitions, we have
\begin{eqnarray*}
  \lefteqn{\sum_{u\in\wt{\cU}}\alpha(u)q_{3,m(j,i)}(P_{\X|U}(\cdot|u))}\\
  & = & E[\Delta_{X^{(i)}}(X^{(i)},\phi_{(j;i)}^*(U,\X^{(\cS_j^c)}))]
\end{eqnarray*}
and
\begin{eqnarray*}
  D_{j,i}
  &\ge& E[\Delta_{X^{(i)}}(X^{(i)},\phi_{(j;i)}(U,\X^{(\cS_j^c)}))]\\
  &\ge& E[\Delta_{X^{(i)}}(X^{(i)},\phi_{(j;i)}^*(U,\X^{(\cS_j^c)}))].
\end{eqnarray*}
Hence, $\phi_{(j;i)}^*$ satisfies the conditions of Theorem
\ref{theorem:manysources:lossy}. Further, Eq.(\ref{eq:proof:multi4}) implies that there
exist a random variable $\wt{U}$ and a joint distribution $P_{\wt{U}\X}$ that satisfy
\begin{eqnarray*}
  \alpha(u)P_{\X|U}(\x^{(\cI_N)}|u) &=& P_{\wt{U}\X}(u,\x^{(\cI_N)})
\end{eqnarray*}
for all $(u,\x^{(\cI_N)})\in\wt{\cU}\times\cX^{(\cI_N)}$. The new joint
distribution preserves the distribution $P_{\X}$
\begin{eqnarray*}
  \sum_{u\in\wt{\cU}}P_{\wt{U}\X}(u,\x^{(\cI_N)})
  &=& \sum_{u\in\wt{\cU}}\alpha(u)P_{\X|U}(\x^{(\cI_N)}|u)\\
  &=& P_{\X}(\x^{(\cI_N)}).
\end{eqnarray*}

This completes the proof of the converse part of Theorem \ref{theorem:manysources:lossy}.
\end{proof}

\subsection{Theorem \ref{theorem:manysources:lossy}: direct part}
\label{sec:proof:multi:achieve}

\begin{proof}~\\
The proof of Theorem \ref{theorem:manysources:lossy} is quite similar to that of Theorem
\ref{theorem:main:lossy}.
Let a set $\D$ of distortion criteria be given, and $P_{U|\X}\in\cP_{CD}(\cU|P_{\X})$.
Fix arbitrary $\gamma,\delta>0$.

\medskip\noindent
\underline{\it Codeword selection: $\varphi_n$}\\
The same way as Theorem \ref{theorem:main:lossy}.

\noindent
\underline{\it Encoding: $\varphi_n$}\\
Almost the same way as Theorem \ref{theorem:main:lossy}.\\
(1) For an input set $\x^{(\cI_N)n}\in\cX^{(\cI_N)n}$ of sequences, the encoder seeks a
    vector $u^n(i)\in\cA_U$ such that $(u^n(i),\x^{(\cI_N)n})\in$
    $T_{U\X}(k_1\delta)$, where $k_1>0$. The selected vector is denoted by
    $u^n(\x^{(\cI_N)n})$.\\
(2) The value assigned to the encoder $\varphi_n(\cdot)$ is the bin index to which
    $u^n(\x^{(\cI_N)n})$ belongs, that is,
    \begin{eqnarray*}
      \varphi_n(\x^{(\cI_N)n}) &=& j, \quad u^n(\x^{(\cI_N)n})\in\cA_U(j).
    \end{eqnarray*}

\noindent
\underline{\it Decoding: $\wh{\varphi}_n^{(j)}$}\\
Almost the same way as Theorem \ref{theorem:main:lossy}.\\
(1) The decoder has access to the indexes $j_U$ received from the encoder $\varphi_n$ and
    the sequence set $\x^{(\cS_j^c)n}\in\cX^{(\cS_j^c)n}$.\\
(2) The decoder seeks a unique vector $u^n\in\cA_U(j_U)$ such that
    $(u^n,\x^{(\cS_j^c)n})\in T_{U\X^{(\cS_j^c)}}(k_{2,j}\delta)$, where $k_{2,j}>0$.
    This vector is denoted by $\wh{u}^n(\x^{(\cS_j^c)n})$.\\
(3) The reconstruction vector $\wh{\x}^{(\cS_j)n}$ is given by
    \begin{eqnarray*}
      \wh{\x}^{(\cS_j)n} &=& \{ \wh{x}^{(i;j)n}|~i\in\cS_j \},\\
      \wh{x}^{(i;j)n}    &=& (\wh{x}_1^{(i;j)},\cdots,\wh{x}_n^{(i;j)}),\\
      \wh{x}_k^{(i;j)}   &=& \phi_{(j;i)}(\wh{u}_k(\x^{(\cS_j^c)n}),\x^{(\cS_j^c)n})
      \quad(k\in\cI_n),
    \end{eqnarray*}
    where $\wh{u}_k(\x^{(\cS_j^c)n})$ is the $k$-th element of $u^n(\x^{(\cS_j^c)n})$.

\noindent
\underline{\it Distortion evaluation: $\wh{\varphi}_n^{(j)}$}\\
  In the same way as Theorem \ref{theorem:main:lossy}, we obtain
  \begin{eqnarray*}
    \lefteqn{\Delta_{X^{(i)}}^n(x^{(i)n},\wh{x}^{(i;j)n})}\nonumber\\
    &=& \frac 1n\hspace{-12mm}\sum_{\hspace{12mm}(u,x^{(\cI_N)})
	\in\cU\times\cX^{(\cI_N)}}\hspace{-13mm}
	N(u,\x^{(\cI_N)}|\wh{u}^n(\x^{(\cS_j^c)n}),\x^{(\cI_N)n})\\
    & & \hspace{10mm}\Delta_{X^{(i)}}(x^{(i)},\phi_{(j;i)}(u,\x^{(\cS_j^c)}))\\
    &\le& \hspace{-12mm}\sum_{\hspace{12mm}(u,\x^{(\cI_N)})\in\cU\times\cX^{(\cI_N)}}
          \hspace{-12mm}(P_{U\X}(u,\x^{(\cI_N)})+k_1\delta)\\
    &   & \hspace{15mm}\Delta_{X^{(i)}}(x^{(i)},\phi_{(j,i)}(u,\x^{(\cS_j^c)}))\\
    &\le& E\left[\Delta_{X^{(i)}}(X^{(i)},\phi_{(j,i)}(U,\X^{(\cS_j^c)}))\right]\\
    &   & \hspace{15mm}+k_1\delta\overline{\Delta}_{X^{(i)}}|\cU\times\cX^{(\cI_N)}|\\
    &\le& D_{j,i}+k_1\delta\overline{\Delta}_{X^{(i)}}|\cU\times\cX^{(\cI_N)}|.
  \end{eqnarray*}
  We denote error probabilities in the encoding/decoding process as $P_e^n$. Then,
  the average distortion can be bounded as
  \begin{eqnarray*}
    \lefteqn{\hspace{-3mm}E\left[\Delta_{X^{(i)}}^n(X^{(i)n},\wh{X}^{(i;j)n})\right]}\\
    &\hspace{-8mm}\le&
      \hspace{-3mm}
      (1-P_e^n)(D_{j,i}+k_1\delta\overline{\Delta}_{X^{(i)}}|\cU\times\cX^{(\cI_N)}|)+
      P_e^n\overline{\Delta}_{X^{(i)}}.
  \end{eqnarray*}
  Since $\delta>0$ is arbitrarily small for a sufficiently large $n$, if $P_e^n$ vanishes
  as $n\to\infty$, we can obtain
  \begin{eqnarray*}
    \limsup_{n\to\infty}E\left[\Delta_{X^{(i)}}^n(X^{(i)n},\wh{X}^{(i;j)n})\right]
    &\le& D_{j,i}.
  \end{eqnarray*}
  
\noindent
\underline{\it Error evaluation: $\varphi_n$}\\
  If there is no $u^n\in\cA_U$ such that
  $(u^n,\x^{(\cI_N)n})\in T_{U\X}(k_1\delta)$, an encoding error has occurred. This
  event is denoted as
  \begin{eqnarray*}
    E_1 &\eqdef& \bigcap_{i=1}^{M_U}
    \left\{(u^n(i),\x^{(\cI_N)n})\notin T_{U\X}(k_1\delta)\right\}.
  \end{eqnarray*}
  Here, let us define
  \begin{eqnarray*}
    E_0 &\eqdef& \{(\x^{(\cI_N)n})\in T_{\X}(k_0\delta)\},
  \end{eqnarray*}
  where $k_0>0$. From Lemma \ref{lemma:typical_prob}, $\Pr\{E_0^c\}\to 0$ as
  $n\to\infty$. Then, in a similar manner to Theorem \ref{theorem:main:lossy}, we have
  \begin{eqnarray*}
    \Pr\{E_1\}
    &\le& \Pr\{E_1\cup E_0^c\}\\
    & = & \Pr\{E_0^c\}+\Pr\{E_{0}\cap E_{1}\},
  \end{eqnarray*}
  \begin{eqnarray*}
    \Pr\{E_0\cap E_1\} &\to& 0 \quad(n\to\infty)
  \end{eqnarray*}
  by setting $M_U$, $k_1$ and $k_0$ as
  \begin{eqnarray*}
    M_U\ge\exp\{n(I(\X;U)+m_1\gamma)\},~m_1>0,
  \end{eqnarray*}
  $m_1\gamma>\epsilon_u=\epsilon_u(k_1\delta,k_0\delta)$ and $k_1<k_0$.\\

\medskip\noindent
\underline{\it Error evaluation: $\wh{\varphi}_n^{(j)}$}\\
  If there is no or more than one $u^n_{(i)}\in\cA_U(j_U)$ such that
  $(u^n_{(i)},\x^{(\cS_j^c)n})\in T_{U\X^{(\cS_j^c)}}(k_2\delta)$, a decoding error is
  declared. This event is classified into two cases.\\
  (1) First case: 
  \[
    (u^n(\x^{(\cI_N)n}),\x^{(\cS_j^c)n})\notin T_{U\X^{(\cS_j^c)}}(k_2\delta).
  \]
  However, this error does not occur by setting $k_2$ as $k_2>k_1|\cX^{(\cS_j)}|$ because
  \[
    (u^n(\x^{(\cI_N)n}),\x^{(\cI_N)n})\in T_{U\X}(k_1\delta)
  \]
  and Lemma \ref{lemma:typical_extend}.\\
  (2) Second case: If there exists $u^n\in\cA_U(j_U)$, $u^n\neq u^n(\x^{(\cI_N)n})$ such
  that $(u^n,\x^{(\cS_j^c)n})\in T_{U\X^{(\cS_j^c)}}(k_2\delta)$. This event is denoted
  as
  \begin{eqnarray*}
    E_2 &\eqdef&
        \hspace{-6mm}\bigcup_{\stackrel{u^n\in\cA_U(j_U)}{u^n\neq u^n(\x^{(\cI_N)n})}}
        \hspace{-7mm}\{(u^n,\x^{(\cS_j^c)n})\in T_{U\X^{(\cS_j^c)}}(k_2\delta)\}.
  \end{eqnarray*}
  Note that if $(\x^{(\cI_N)n})\in T_{\X}(k_0\delta)$ then
  \begin{eqnarray*}
    \x^{(\cS_j^c)n} &\in& T_{\X^{(\cS_j^c)}}(k_0\delta|\cX^{(\cS_j)}|).
  \end{eqnarray*}
  Therefore, we have
  \begin{eqnarray*}
    \Pr\{E_2\}
    &\le& \Pr\{E_2\cup E_0^c\}\\
    & = & \Pr\{E_0^c\}+\Pr\{E_0\cap E_2\},\\
    \Pr\{E_0\cap E_2\}
    &\to& 0\quad(n\to\infty)
  \end{eqnarray*}
  in a similar manner to Theorem \ref{theorem:main:lossy} by setting $L_U$, $k_2$ as
  \begin{eqnarray*}
    L_U\le\exp\{n(I(\X^{(\cS_j^c)};U)-l_{1j}\gamma)\},~l_1>0,
  \end{eqnarray*}
  $l_{1j}\gamma>\epsilon=\epsilon(k_0\delta|\cX^{(\cS_j)}|,k_2\delta)$ and
  $k_0|\cX^{(\cS_j)}|<k_2$.\\

\medskip\noindent
\underline{\it Rate evaluation: $\varphi_n$}\\
  The encoder sends the indexes of the bin using
  \begin{eqnarray*}
    \lefteqn{\hspace{-3mm}R = \frac 1n\log N_U}\nonumber\\
    &\hspace{-3mm} = & \frac 1n\log\frac{M_U}{L_U}\\
    &\hspace{-3mm}\ge& I(\X;U)+m_{1}\gamma-\min_{j\in\cI_M}\{I(\X^{(\cS_j^c)};U)
                       -l_{1j}\gamma\}\\
    &\hspace{-3mm} = & \max_{j\in\cS_j}\{I(\X^{(\cS_j)};U|\X^{(\cS_j^c)})+l_{1j}\gamma\}
		       +m_{1}\gamma
  \end{eqnarray*}
  bits per letter. Since $\gamma>0$ is arbitrary, we obtain the coding rate as
  $\max_{j\in\cS_j}I(\X^{(\cS_j)};U|\X^{(\cS_j^c)})$.

This completes the proof of Theorem \ref{theorem:manysources:lossy}.
\end{proof}

\section*{Acknowledgments}
The authors wish to thank Prof. Ryutaroh Matsumoto of Tokyo Institute of Technology for
his support. The authors also thank Dr. Yoshinobu Tonomura, Dr. Hiromi Nakaiwa, Dr. Shoji
Makino, Dr. Junji Yamato and Dr. Kunio Kashino of NTT Communication Science Laboratories
for their help. Lastly, the authors thank the associate editor and the anonymous
reviewers for their constructive remarks and suggestions, which led to the improvement of
this work.

\bibliographystyle{bib/ieicetr}
\bibliography{bib/IEEEabrv,bib/defs,bib/it}

\profile*{Akisato Kimura}{
received B.E., M.E. and D.E. degrees from Tokyo Institute of Technology in 1998, 2000 and
2007, respectively. In 2000, he joined NTT Communication Science Laboratories, Nippon
Telegraph and Telephone Corporation, Japan, where he is currently a research scientist in
Media Information Laboratory. He has been working on multimedia information retrieval,
perceptual image processing and multiterminal information theory. His research interests
are in the areas of pattern recognition, computer vision and information theory.
}
\profile*{Tomohiko Uyematsu}{
received B.E., M.E. and D.E. degrees from Tokyo Institute of Technology in 1982, 1984 and
1988, respectively. From 1984 to 1992, he was with the Department of Electrical and
Electronics Engineering of Tokyo Institute of Technology, first as a research associate,
next as a lecturer, and lastly as an associate professor. From 1992 to 1997, he was with
School of Information Science of Japan Advanced Institute of Science and Technology as an
associate professor, and currently he is a professor in the Department of Communications
and Integrated Systems. In 1992 and 1996, he was a visiting researcher at the Centre
National de la Recherche Scientifique, France and Delft University of Technology,
Netherlands, respectively. He received the Shinohara Memorial Young Engineer Award in
1989, and the Best Paper Award in 1993, 1996, 2002 and 2007 all from IEICE. His current
research interests are in the areas of information theory, especially Shannon theory and
multiterminal information theory. Dr. Uyematsu is a senior member of IEEE.
}

\end{document}